\documentclass[11pt]{article}
\usepackage[english]{babel}

\usepackage[letterpaper,top=4cm,bottom=4cm,left=3cm,right=3cm,marginparwidth=1.75cm]{geometry}
\usepackage[square,sort,comma,numbers]{natbib}
\usepackage{amsmath}
\usepackage{wasysym}
\usepackage{amssymb}

\usepackage{amsthm}
\usepackage{program}
\usepackage{graphicx}
\usepackage[colorlinks=true, allcolors=blue]{hyperref}
\usepackage{blindtext}
\usepackage{float}
\usepackage{subfig}
\usepackage{fancybox,graphicx}
\usepackage{subfig}

\usepackage{color}
\usepackage{authblk}
\usepackage{wrapfig}

\usepackage[titletoc,title]{appendix}
\usepackage{enumitem}
\usepackage{booktabs}
\usepackage{mathtools}

\usepackage[compatibility=false]{caption}

\title{The RISTRETTO simulator:\\ Exoplanet reflected spectra} 
 
\author[1]{M. Bugatti}
\author[1]{C. Lovis}
\author[1]{F. Pepe}
\author[1]{N. Blind}
\author[1]{N. Billot}
\author[1]{B. Chazelas}
\author[2,3]{M. Turbet}

\affil[1]{D\'epartement d'Astronomie, Universit\'e de Gen\`eve, Chemin Pegasi 51, CH-1290 Versoix, Switzerland}
\affil[2]{Laboratoire de M\'et\'eorologie Dynamique, IPSL, CNRS, Sorbonne Universit\'e, 4 place Jussieu, F-75252 Paris Cedex 05, France}
\affil[3]{Laboratoire d'astrophysique de Bordeaux, Univ. Bordeaux, CNRS, B18N, allée Geoffroy Saint-Hilaire, 33615 Pessac, France}

\begin{document}
\maketitle
\begin{abstract}
\noindent The upcoming Ristretto spectrograph is dedicated to the detection and analysis of exoplanetary atmospheres, with a primary focus on the temperate rocky world Proxima b. This scientific endeavor relies on the interplay of a high-contrast adaptive optics (AO) system and a high-resolution echelle spectrograph. In this work, I present a comprehensive simulation of Ristretto's output spectra, employing the Python package Pyechelle. Starting from realistic spectra of both exoplanets and their host stars, I generate synthetic 2D spectra to closely resemble those that will be produced by Ristretto itself. These synthetic spectra are subsequently treated as authentic data and therefore analyzed. These simulations facilitate not only the investigation of potential exoplanetary atmospheres but also an in-depth assessment of the inherent capabilities and limitations of the Ristretto spectrograph.\\\\

\noindent Keywords: High-resolution spectrograph, High spatial resolution, simulations, reflection spectroscopy.  
\end{abstract}

\section{Introduction}
Detecting and characterizing Earth-like exoplanets represents one of the most challenging goals in modern exoplanet research. RISTRETTO, a high-contrast and high-resolution spectrograph being developed at the University of Geneva, aims to address these challenges by studying the reflected light spectra of nearby exoplanets \citep{lovis2022ristretto}. Planned as a visitor instrument for ESO's VLT, RISTRETTO is composed of two main parts: the front-end, which includes an eXtreme Adaptive Optics (XAO) system and a coronagraph/apodizer to minimize the star's brightness, and the back-end, which consists of a high-resolution spectrograph, a fiber link, and an integral-field unit (IFU) to isolate the planet's signal.\\
Detecting Earth-like exoplanets is particularly challenging due to several factors. Firstly, the small brightness ratios between the planet and its host star result in a low signal-to-noise ratio, making it difficult to discern the faint signature of an exoplanet against the overwhelming glare of its star. Secondly, the habitable zones of such exoplanets, where conditions are conducive to liquid water, are often close to the star, complicating the angular separation of the planet from its host. 
Lastly, the majority of exoplanets does not perform transit, and therefore necessitate characterization through reflection and thermal emission spectra.\\
\noindent While so far there is no space mission dedicated to studying the atmospheres of small non-transiting exoplanets, many ongoing and future ground-based projects aim to study exoplanets' reflected light. Examples include ANDES, and PCS on the Extremely Large Telescope (ELT) \citep{palle2023ground}; IRIS and MICHI on the Thirty Meter Telescope (TMT); and GMTIFS and GMagAO-X + IFS on the Giant Magellan Telescope (GMT).\\
The difficulties in detecting Earth-like planets in the habitable zone have led recent investigations to focus on smaller stars, particularly M-dwarfs. Their diminished stellar radius and luminosity result in closer conventional habitable zones, offering more favorable contrast ratios and transit probabilities \citep{charbonneau2007dynamics}. Proxima Centauri b, orbiting the M-dwarf Proxima Centauri - the closest star to the Sun - presents an enticing opportunity for studying a potentially Earth-like exoplanet. Located within the star's habitable zone \citep{faria2022candidate}, Proxima b's proximity to Earth makes it an accessible target for observation and characterization studies through reflection spectra.\\
In this paper, we focus on simulating the detection of Proxima b by RISTRETTO. These simulations are essential for modeling and predicting the behavior of the spectrograph under various conditions, aiding in the optimization of the instrument's working conditions and data analysis. By simulating spectra, potential issues such as instrument limits can be identified and corrected before actual observations, saving valuable telescope time. Additionally, simulations assist in developing data reduction and analysis pipelines, ensuring they are robust and accurate when applied to real observational data.\\
The paper is divided as follows: the second section presents RISTRETTO specifics and data acquisition; section three focuses on the simulations, from the synthetic input spectra to the actual planet detection; section four discusses the conclusions drawn from our work and outlines potential avenues for future research.

\section{RISTRETTO}
\subsection{Instrument specifics}
RISTRETTO combines a high-contrast AO system working at the diffraction
limit of the telescope to a high-resolution spectrograph, via a 7-spaxel integral-field unit (IFU) feeding singlemode fibers \citep{chazelas2020ristretto}. The calibration system is composed of a Neon Uranium lamp which is used for the wavelength calibration and a Fabry Perot Interferometer to measure the drift of the instrument during the night (simultaneous reference). Here below are listed the properties of the spectrograph and of the feeding fibers \citep{chazelas2020ristretto}:
\begin{enumerate}
\itemsep 0em
    \item Spectral resolution $=R_s>=130000$, goal $150000$
    \item Band: $\lambda_{min}=620$ nm, $\lambda_{max}=840$ nm.
    \item The detector is the CCD 23184 from E2V-teledyne \citep {e2v-datasheet}.
    \item The spatial projection on the sky between two spaxels is about $37$ milliarcseconds.
    \item The average total spectrograph transmission: $\sim 7.7\%$, which is computed through simulations performed by Nicolas Blind taking into account the optical systems of the spectrograph. It includes the following contributions:
    \begin{enumerate}
        \item Average Transmission of the atmosphere on the RISTRETTO science band at the Zenith: $96.6 \%$
        \item Aluminum coating transmission $61.2 \%$
        \item Front-end transmission: $68.1 \%$
        \item Adaptive Optics + Coronograph transmission: it depends on the position on the sky of the detected object and it will be discussed in detail in subsection \ref{subsec3.4}
        \item Fiber link efficiency: $89.1 \%$
        \item Average spectrograph efficiency: $43.9 \%$
    \end{enumerate}
\end{enumerate}
\label{sub2.1}
\subsection{RISTRETTO observations}
\begin{wrapfigure}{r}{0.5\textwidth}
\includegraphics[width=0.5\textwidth]{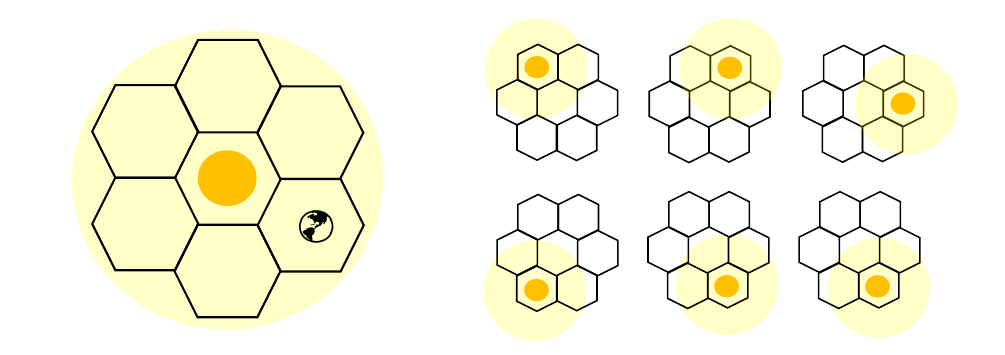}
    \caption{Left side: RISTRETTO first long-time exposure. Right side: RISTRETTO $6$ short-time exposures with the star centered in the off-axis spaxels.}
    \label{1}
\end{wrapfigure}

In the manner of real future observations, RISTRETTO's central spaxel will point at the center of the host star, ensuring that the six off-axis spaxels capture the star's halo, with one of them also containing the exoplanet's reflected light. Initially, a long exposure is performed using a filter in the central spaxel to prevent saturation. Subsequently, six short exposures are taken with the host star centered in each of the six off-axis spaxels. This step is crucial for later modeling of the star's halo spectrum, mitigating the effects of the variable Point Spread Function (PSF) for each off-axis spaxel. A visual diagram of the 1+6 exposures is shown in Figure \ref{1}.

\section{Simulating RISTRETTO observations}
In our simulations, we follow a step-by-step process to mimic real observations and ensure that the simulated spectra accurately reflect the complexities of real observational conditions:
\begin{enumerate}[label=\arabic*.]

    \item \textbf{Generate synthetic spectra:} we craft synthetic spectra for the host star, its surrounding halo, and the orbiting exoplanet.

    \item \textbf{Design observation parameters:} after assuming some orbital parameters governing the trajectory of the target exoplanet, we select the epochs of observation.
    
    \item \textbf{Compute the radial velocities:} we precisely compute the radial velocities for each exposure, and we apply the corresponding Doppler shift to each spectrum.
    
    \item \textbf{Incorporate the efficiency and fiber coupling functions:} we integrate the spectrograph efficiency and fiber coupling functions into each spectrum to account for instrumental artifacts and refine the fidelity of the simulated data.
    
    \item \textbf{Generate 2D spectra:} we use Pyechelle, a Python package, to generate 2D spectra, incorporating Point Spread Functions (PSF), noise profiles, and realistic variations in Earth's atmospheric conditions.
    
    \item \textbf{Extract 1D spectrum:} we employ the Optimal extraction technique to extract a 1D spectrum from the 2D raw frame generated with Pyechelle.
    
    \item \textbf{Identify planet signal:} we identify and analyze the planetary signal, advancing our understanding of data analysis and instrument limitations.
\end{enumerate}
\label{sec3}
\subsection{Generate synthetic spectra}
\begin{figure}[ht]
    \centering
\includegraphics[width=0.5\textwidth]{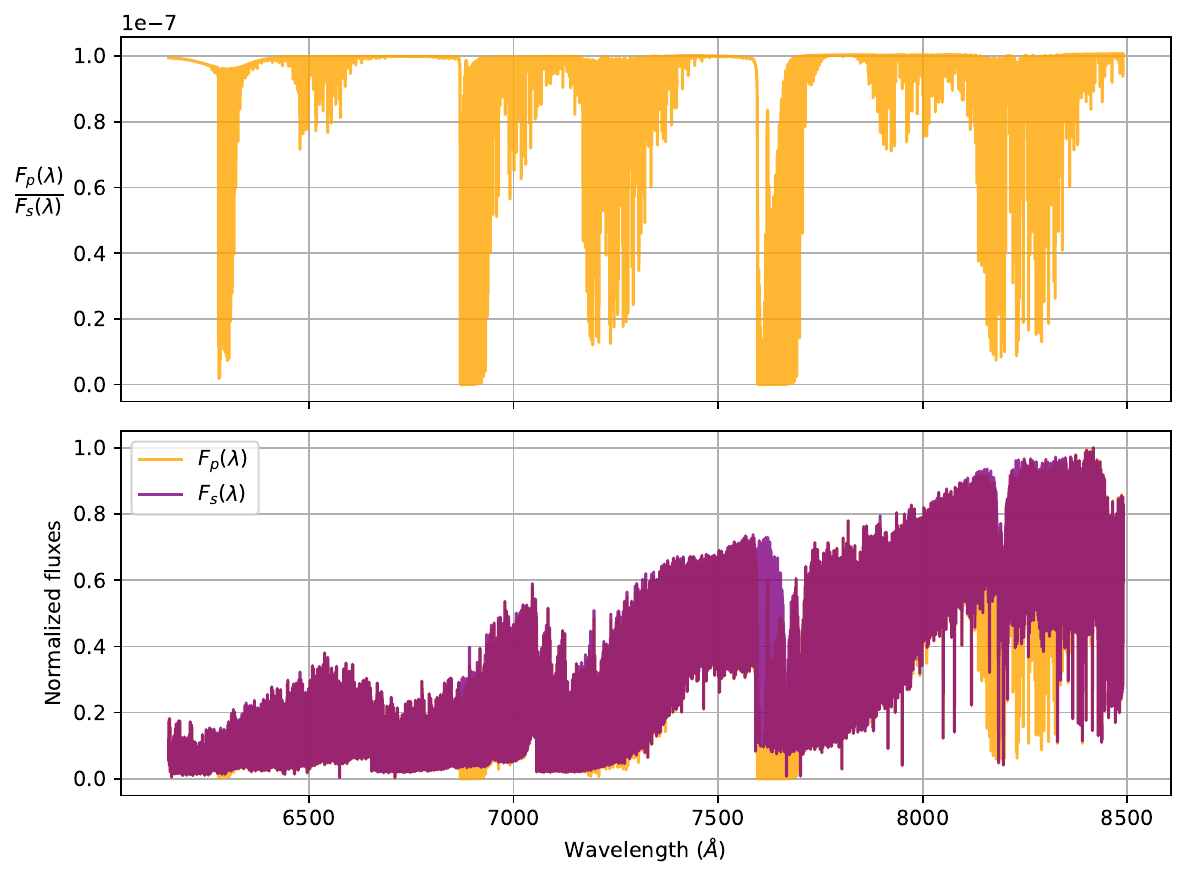}
\caption{Upper plot: $F_p/F_s$, lower plot: Normalized $F_p$ and $F_s$. $F_p$ is expressed at a phase of $90$°. All spectra have a resolution of $\sim 500'000$.}
    \label{3}
\end{figure}

To compute the input spectra we used the
data of Table \ref{table1},
\begin{table}[ht]
    \centering
    \caption{Proxima and Proxima b properties. \textbf{References:} 1- \citep{collaboration2020vizier}, 2- \citep{husser2013new}, 3- \citep{faria2022candidate}, 4- \citep{boyajian2012stellar}}
    \begin{tabular}{lll}
        \toprule
        Parameters Proxima Centauri & Value & ref \\
        \midrule
        Mass & $M_s=0.122$ $M_{\sun}$ & 1\\
        Radius & $R_{\star}=0.141$ $R_{\sun}$ & 4\\
        Distance from Earth & $d=268391.77$ AU & 1 \\
        Right Ascension&  $14 : 29 : 42.94613 2$ & 1 \\
        Declination & $ -62:40: 46.16468 $ & 1 \\

        PHOENIX Model & & \\
        & $T_{eff}=3000$ K & 2\\
        & $\log{(g)}=5 $ & 2\\
        & $[Fe/H]=0$ & 2\\
        & $[\alpha/M]=0$ & 2\\\\
        \toprule
        Parameters Proxima b & Value &\\
        \midrule
        Minimum Mass & $ M_p \sin i=1.07$ $M_{\oplus}$ & 3\\
        Orbital Period & $P=11.187$ days & 3\\
        Semi-Major Axis &  $a=0.0486$ AU & 3\\
        Eccentricity & $ e=0.02$ & 3\\
        Argument of Periastron & $\omega=3.3$° & 3\\
        Radial velocity amplitude & $K_s=1.24$ m/s & 3\\ 

        
    \end{tabular}
    \label{table1}
\end{table}
\noindent Specifically, to produce the star spectrum we used the Python package Expecto \citep{expecto_cite}, using as input the effective temperature $T_{eff}$ of the star and its surface gravity $\log{(g)}$. Expecto uses the PHOENIX model \citep{husser2013new} to create the spectrum \citep{Earl2023-kz} emitted on the surface of the star in units of $erg/s/cm^2/cm$.\\ 
\noindent Concerning the contribution from the planet Proxima b, we generated high-resolution (R=500,000) albedo spectra of the planet by coupling the outputs of a 3D Global Climate Model \citep{turbet2016habitability} with the radiative transfer code PICASO \citep{batalha2019exoplanet}. We assumed here an Earth-like atmosphere with a 1bar, N2-O2-dominated atmosphere, with 400ppm of CO2, and a surface fully covered with water. Water vapor and clouds are variable and calculated in a self-consistent way by the 3D model. Molecular absorptions from all four molecules are taken into account in our radiative transfer calculations, as well as Rayleigh and Mie scattering.\\ 
\noindent From now on, we will call the theoretical star spectrum $F_s(\lambda)$, the theoretical planet spectrum $F_p(\lambda)$, and the albedo spectrum $F_p(\lambda)/F_s(\lambda)$. All of them are shown in Figure \ref{3}. 



\subsection{Design observation parameters}
\begin{figure}[ht]
    \centering
\includegraphics[width=\textwidth]{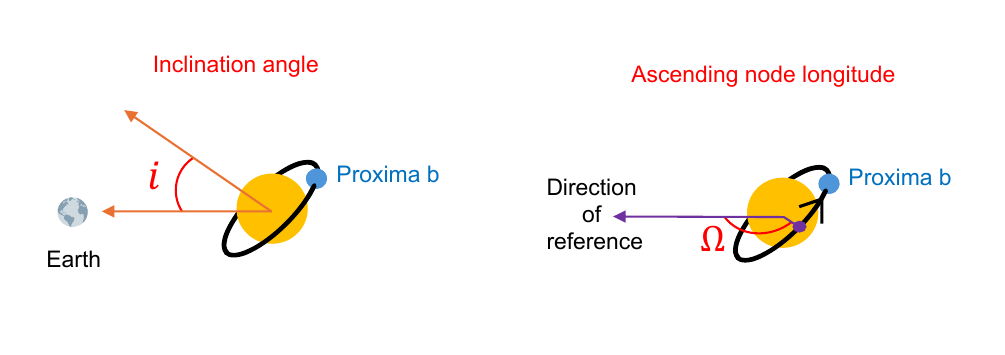}
\caption{Inclination angle and ascending node longitude representation}
    \label{2}
\end{figure}

When detecting an exoplanet using the radial velocity method, such as Proxima b, two critical unknown parameters of the exoplanet's orbit are the inclination angle ($i$) and the longitude of the ascending node ($\Omega$) of the orbit plane \citep{hatzes2016radial}. These parameters scheme is illustrated in Figure \ref{2}.

\begin{figure}[ht]
    \centering
\includegraphics[width=0.8\textwidth]{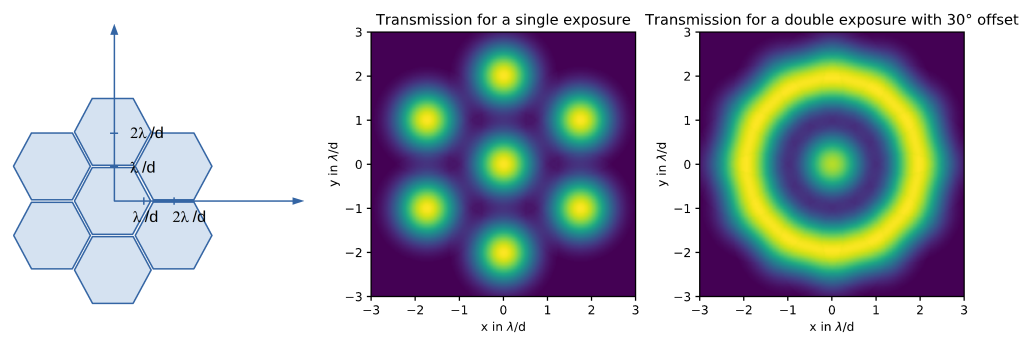}
    \caption{Left panel: disposition of the input fibers on the sky. Central panel: one exposure transmission. Right panel: transmission after $2$ exposures with a $30^\circ$ offset}
    \label{9}
\end{figure}

\noindent Understanding these parameters is essential for accurately determining the exoplanet's orbit as observed from Earth. Specifically, knowing the ascending node longitude ($\Omega$) helps pinpoint the exact spaxel where the exoplanet will appear at its maximum elongation from the host star. Meanwhile, the inclination angle ($i$) is necessary to accurately calculate the exoplanet's orbital radial velocity.
In our simulations, we assume prior knowledge of the longitude of the ascending node. 
If $\Omega$ cannot be determined beforehand, RISTRETTO will mitigate this uncertainty by taking two consecutive exposures, each rotated $30$ degrees relative to the other. This approach ensures comprehensive coverage of the star's surrounding area, enhancing the likelihood of detecting the exoplanet as shown in Figure \ref{9}.\\

\noindent To simulate the orbit and the radial velocities of Proxima b we took the data found in Table \ref{table1}, and we assumed two hypothetical values for the unknown parameters:
\begin{itemize}
    \item The inclination of the orbit $i=60$° \item The longitude of the ascending node $\Omega= 0$°
\end{itemize}
\noindent The orbit of Proxima b on the sky can be seen on the right side of Figure \ref{6} - in black. It was computed by converting orbital elements to Cartesian coordinates (in milliarcseconds) by iteratively solving Kepler's equation for the eccentric anomaly using the Newton-Raphson's method until convergence \citep{schwarz2017keplerian}, \citep{roblox_keplerian}.\\
We simulated 11 nights of observations, choosing intervals as close as possible to the period $P$. This approach helps maximize the planet's orbital radial velocity signal, making it easier to distinguish from the star's signal.
For each night, we assumed $9$ exposures of $1$ hour each to detect the planet in an off-axis spaxel (left side of Figure \ref{1}). After each 1-hour exposure, we assumed $6$ additional exposures of $360$ seconds each, with the star centered in each off-axis spaxel (right side of Figure \ref{1}).


\subsection{Compute the radial velocities:}
\noindent We calculated the radial velocities of both the planet and the star throughout their orbits to adjust the Doppler shift in the input spectra for each exposure. We considered the following three types of radial velocities:
\begin{enumerate}
    \item \textbf{Systemic radial velocity} - 
    it is the velocity component along the line of sight between the observer and the exosystem star + planet(s), measured in the observer's rest frame. It includes any motion of the star due to its intrinsic properties (e.g., its orbit around the galactic center). It has a fixed value and for the Proxima system is equal to  $-21700$ m/s. \citep{lovis2017atmospheric}.
    \item \textbf{Orbital radial velocity} - it refers to the radial velocity component of the astronomical object caused by its orbital motion around the center of mass of the system. We used the following formula to calculate it \citep{clubb2008detailed}: 
\begin{equation}
V_r=K\cdot[\cos(\theta+w)+e\cos(w)]\quad \text{where} \quad \theta=2 \pi  t / P
\end{equation}
  Where $V_r$ is the orbital radial velocity, $K$ is the radial velocity semi‐amplitude of the orbiting object, $\omega$ is the argument of periastron, $e$ the eccentricity, $P$ the orbital period around the center of mass, and $t$ the time epoch at which the observer is looking at the orbiting object. $K_s$ of Table \ref{table1} is the semi‐amplitude of the star, from which the minimum planet mass $M_p\sin i$ is computed. While $K_p$, namely the radial velocity semi-amplitude of the orbiting planet, is unknown.
Using the data of Table \ref{table1} and $i=60^{\circ}$, the computed orbital radial velocities for the planet and the star-planet separation are shown in Figure  \ref{5}.
In red are represented the observation exposures, which are the closest to the maximum elongation of the planet.
\begin{figure}[ht]
    \centering
\includegraphics[width=0.9\textwidth]{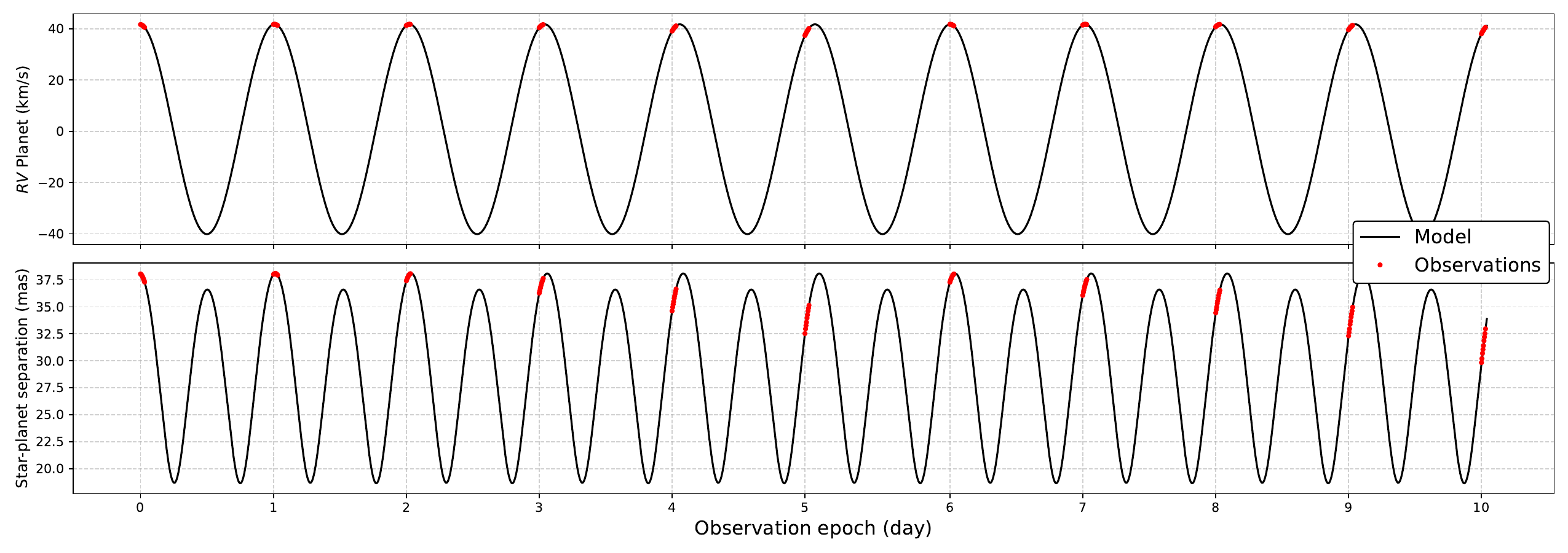}
\caption{Orbital radial velocity of star and planet}
    \label{5}
\end{figure} 
\item \textbf{Barycentric radial velocity}

\begin{minipage}{.45\textwidth}
\vspace{-8pt}
It refers to the velocity of an astronomical object relative to the barycenter of the solar system. 
Barycentric radial velocity includes the motions of the observer, such as the Earth's motion around the Sun and the Earth's rotation. This velocity is the same for both Proxima and Proxima b and depends on the date and time of the observations.
\end{minipage}
    \begin{minipage}{.40\textwidth}
    \vspace{-15pt}
    \includegraphics[scale=0.35]{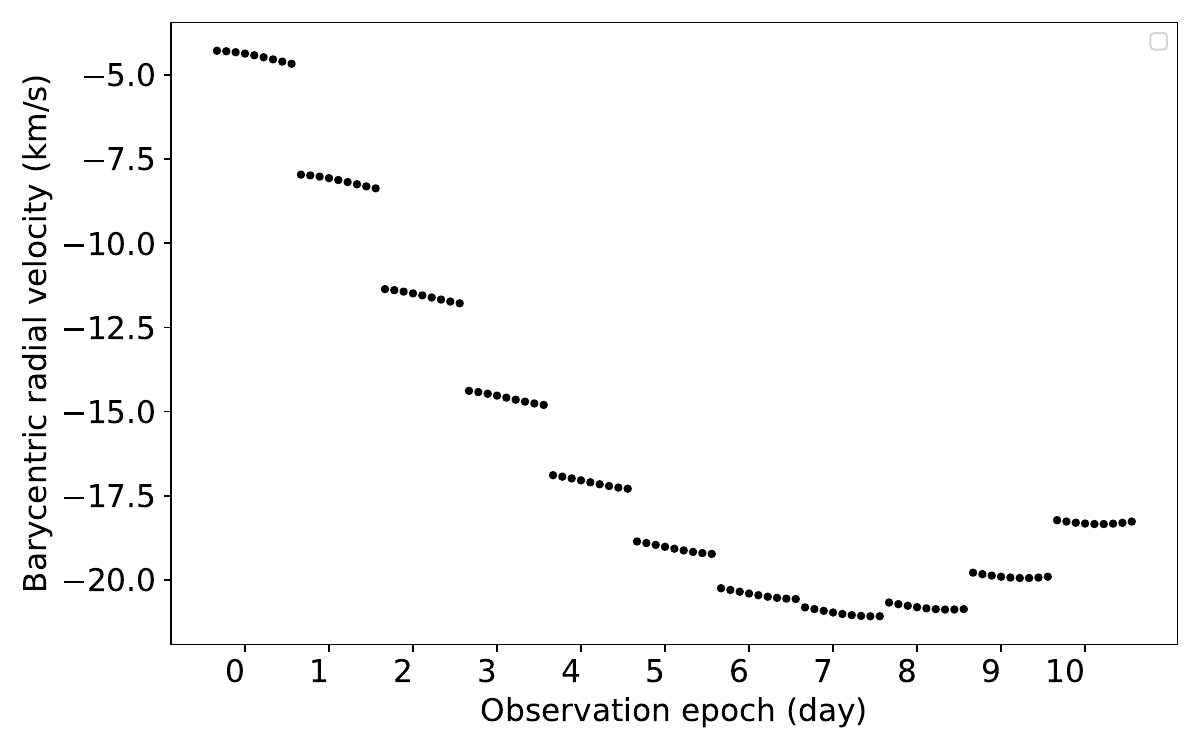}
      \label{4}
      
    \end{minipage}

\vspace{-10pt}
We computed this latter radial velocity using the python package Barycorrpy \citep{Shubham2021-nh}, which has a specific function (\verb|get_BC_vel|) that given a time of observation, a target and a place of observation (VLT, Paranal Observatory, Chile) automatically calculates the Barycentric radial velocity. Figure \ref{4} represents the Barycentric radial velocities, for both the planet and the star, during $11$ days of observations in June-July-August 2022. In fact, we assumed to observe Proxima b from VLT, where RISTRETTO is planned to be used as a visitor instrument.  
There, in the southern hemisphere, and according to his position in the sky (Table \ref{table1}), Proxima Centauri is best observed during the austral winter months, namely June, July, and August as the star will be visible for longer durations and at higher altitudes.\\
\end{enumerate}

\subsection{Incorporate the efficiency and fiber coupling functions:}
As efficiency, we used the components specified in subsection \ref{sub2.1}: \(a \cdot b \cdot c \cdot d \cdot e \cdot f = 15.7\% \cdot d\). To estimate \(d\) accurately, we used data simulated by Nicolas Blind in \citep{blind2022ristretto}, which represent the combined efficiency of both the coronograph and the Adaptive Optics system. The simulations were conducted for 7 different fibers, 5 wavelengths \([620, 675, 730, 785, 840] \text{ nm}\), and various positions in the sky of the observed object (values ranging from \(-60\) to \(+60\) milliarcseconds on both the \(x\) and \(y\) axes, with intervals of 2 milliarcseconds).\\
We interpolated over the 5 wavelengths and the \(60 \times 60\) values of \(x\) and \(y\) in milliarcseconds for intermediate points corresponding to our observation positions.\\
\begin{figure}[ht]
\centering
\includegraphics[width=\textwidth]{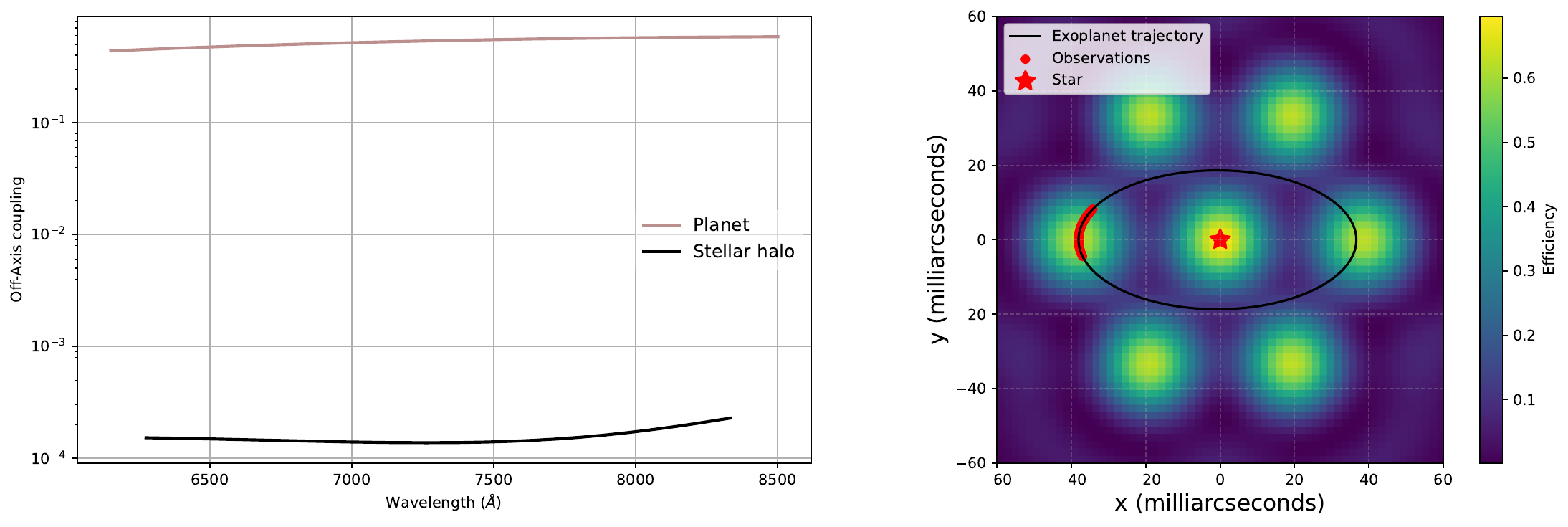}
\caption{Left side: it shows how the coupling depends on wavelength when the star is in the central spaxel and the planet is in an off-axis spaxel. Right side: it illustrates the coupling of $7$ fibers at a wavelength of $840$ nm, with the trajectory of the exoplanet overlaid on top.}
    \label{6}
\end{figure}
\noindent The left side of Figure \ref{6} illustrates how the coupling of an off-axis spaxel varies with wavelength. The star is positioned in the central spaxel and the planet at the center of the off-axis spaxel.
The right side of Figure \ref{6} shows an example of coupling at \(\lambda = 840\) nm for the 7 fibers, based on the position of the astronomical object in the sky in milliarcseconds ($mas$).

\label{subsec3.4}
\subsection{Generate 2D spectra}
\begin{figure}[ht]
\centering
\includegraphics[width=\textwidth]{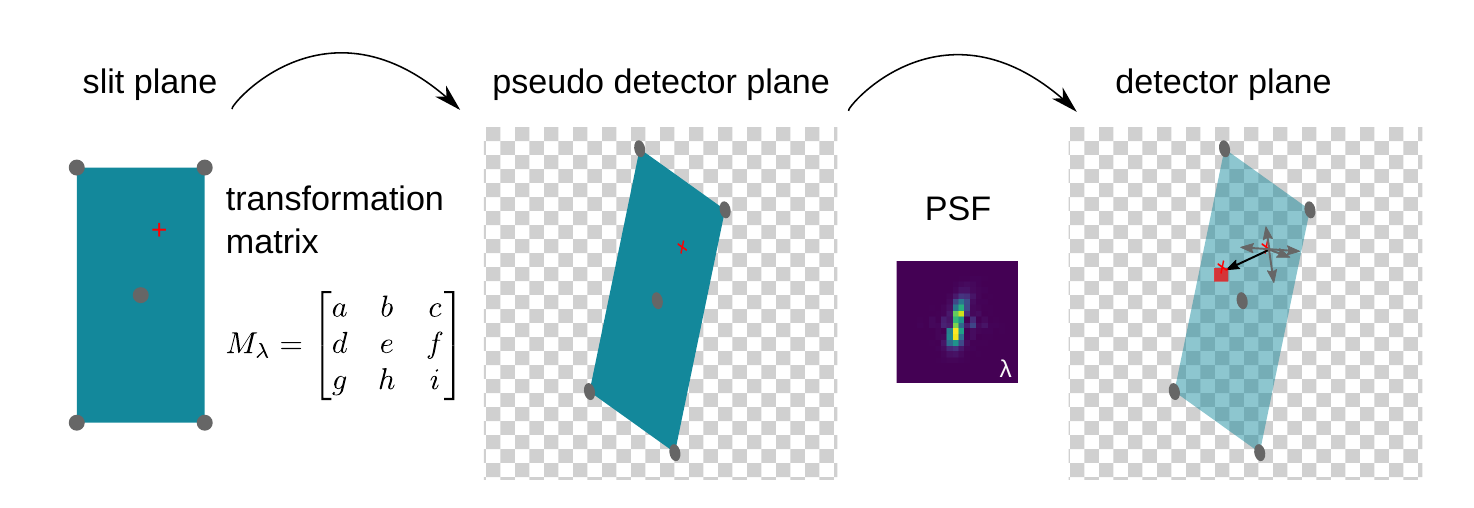}
\caption{Pyechelle workflow}
    \label{11}
\end{figure}

To simulate RISTRETTO spectra, we utilized the Python package PyEchelle, specifically designed for generating realistic 2D cross-dispersed echelle spectra. PyEchelle incorporates optical aberrations and can simulate atmospheric telluric transmission, along with factors like readout noise, photon noise, and bias to enhance realism. It operates on the principle that any spectrograph can be represented using wavelength-dependent transformation matrices and point spread functions to describe its optics \citep{sturmer2018echelle++}. For clarity, transformation matrices are matrices that transform one vector into another through matrix multiplication, preserving points, straight lines, and planes, though not necessarily distances or angles. \\
In PyEchelle, these matrices and point spread functions are derived from the RISTRETTO model in ZEMAX, an optical modeling software \citep{zemax}. As the geometric transformations associated with the matrices vary smoothly across an echelle order, they are interpolated in PyEchelle for any intermediate wavelength. The same applies to the wavelength-dependent point spread functions, which slowly vary across an echelle order, allowing for interpolation at intermediate wavelengths \citep{sturmer2018echelle++}.
The scheme of Figure \ref{11} shows the workflow of Pyechelle.\\

\noindent Before using synthetic spectra as input for PyEchelle, we needed to convert the units from \( \text{erg/s/cm}^2/\text{cm} \) to photons/s/cm, which PyEchelle uses to generate photons and produce 2D spectra. Let \( F_{\text{erg}}(\lambda) \) denote the flux spectrum of Proxima at its surface, computed with Expecto and expressed in \( \text{erg/s/cm}^2/\text{cm} \). To convert \( F_{\text{erg}}(\lambda) \) to \( F_{\text{ph}}(\lambda) \) in photons/s/cm units at the telescope's entrance, we performed the calculation:

\[
F_{\text{ph}}(\lambda) = F_{\text{erg}}(\lambda) \cdot \frac{R_{\star}^2}{d^2} \cdot \frac{\lambda}{hc} \cdot A_{\text{telescope}}
\]

\noindent Here, \( h \) is Planck's constant, \( c \) is the speed of light, \( A_{\text{telescope}} \) is the collective area of the telescope (49.3 \( \text{m}^2 \) for the VLT), and \( R_{\star} \) and \( d \) are specified in Table \ref{table1}.\\
PyEchelle also requires separate inputs for the CCD bias and read-out noise, as well as Earth atmosphere conditions for generating telluric lines. For the CCD, we assumed perfect pixels without Charge Transfer Inefficiency or Dark current effects, as cooling/temperature effects are not included in the CCD model in PyEchelle. 
For the bias, we decided to use an arbitrary value of $250$ photons. Using an estimated bias has minimal impact as it merely adds a constant value to each pixel. In our simulations, we set this value to 250 photons. For the atmosphere conditions we set a random airmass between 1 and 2.\\
Figure 7 displays a simulated 2D raw frame generated with PyEchelle. Each order contains seven lines, with each line representing the 2D spectrum of a spaxel. The wavelength solution varies significantly across spaxels within the same order due to physical limits on the vertical distance between fiber core centers. The cladding of mono-mode fibers prevents achieving the required vertical distance, resulting in a horizontal shift of fibers and a considerable difference in the wavelength solution from one spaxel to another.

\begin{figure}[ht]
\centering
\includegraphics[width=0.8\textwidth]{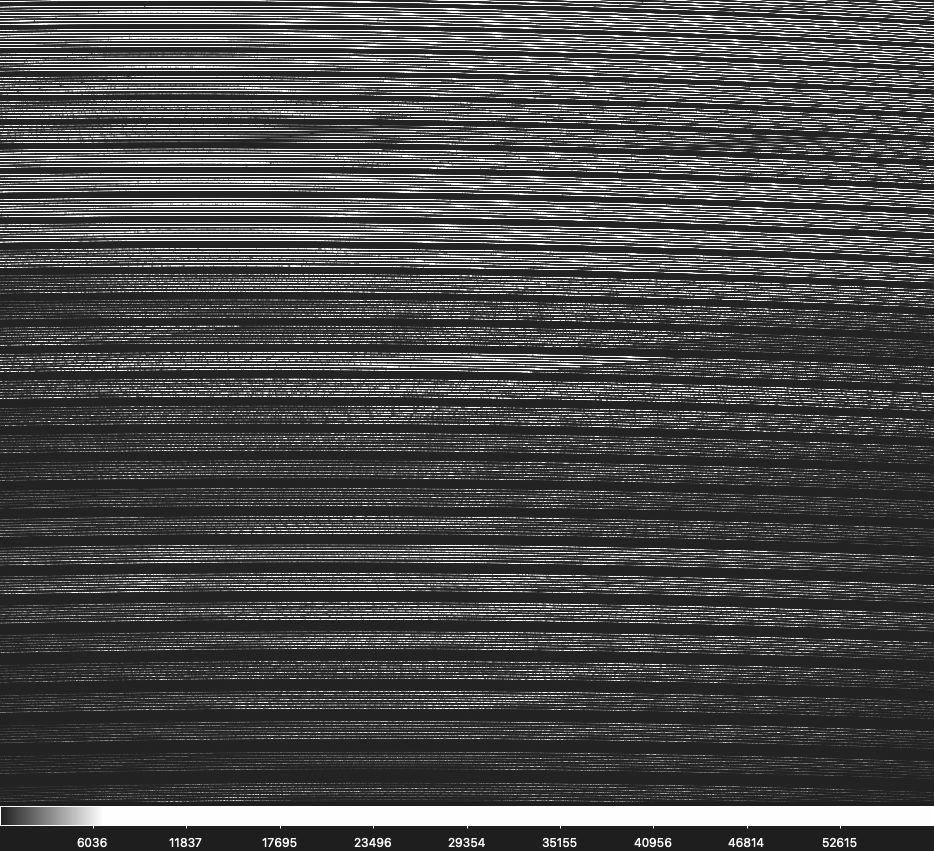}
\caption{Simulated 2D raw frame with the RISTRETTO model}
    \label{7}
\end{figure}

\subsection{Extract 1D spectrum}
To extract the 1D spectrum from the raw frame we employed the Optimal Extraction technique shown in \citep{horne1986optimal}.
The optimal extraction method takes into account the irregular profile of echelle orders, compensating for variations in instrumental response and optimizing signal-to-noise ratios.\\
In our specific case, we are dealing with perfect pixels, without Charge Transfer Inefficiency, nor Dark current effects, which we did not include in the simulations. 
Our analysis is therefore a simplification of the whole Optimal Extraction procedure performed in realistic conditions. \\
By construction, on the CCD, the wavelength direction coincides with the detector $X$ direction and the cross-dispersion direction with the detector $Y$ direction.
We show the steps that we used in the simulation to extract a one-dimensional spectrum.

\begin{figure}[ht]
\centering
\includegraphics[width=\textwidth]{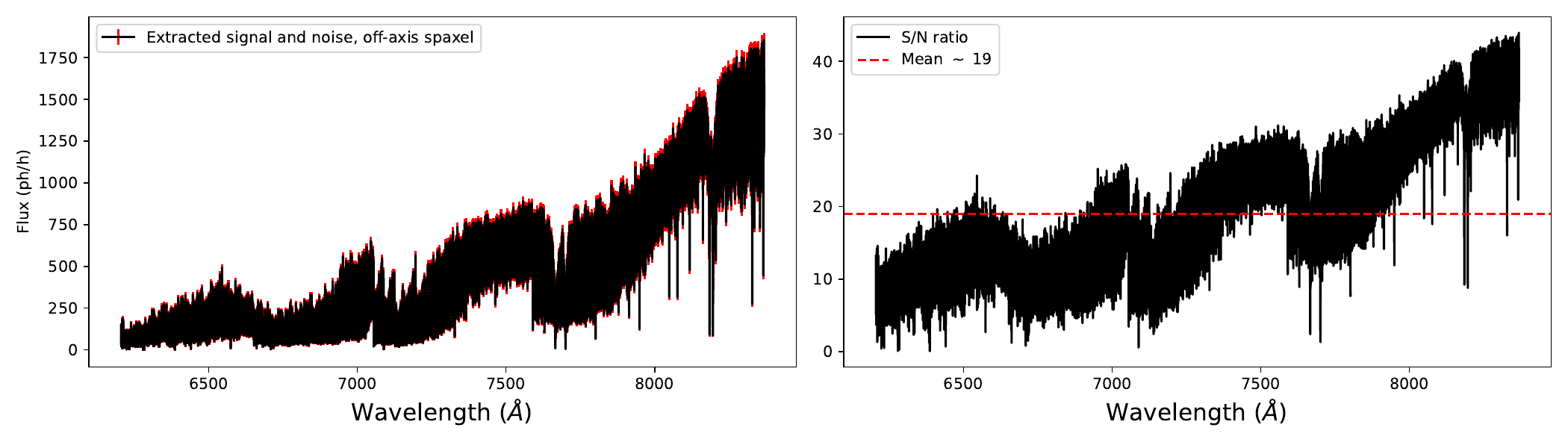}
\caption{Left panel: extracted 1D spectrum of star's halo + planet, for 1 hour exposure. Right panel: corresponding signal-to-noise ratio}
    \label{12}
\end{figure}

\begin{enumerate}
    \item We use the pre-existing model from ZEMAX to trace the order profiles, avoiding the need to compute them ourselves from a flat-field image. Order profiles denote the position of the order center in pixels along the wavelength direction. Similarly, we use the wavelength solution already available in the Ristretto model, represented as a function $\lambda(x)$, with $x$ being the horizontal pixel position.
    \item We generate the Master Flat-Field frame by simulating 10 flat-field exposures using a constant photon source. After subtracting the bias, we average the $10$ images to create the Master Flat Field.
    \item From the Master Flat Field, we compute the normalized trace profile \(P_{xy}\) by taking a 10-pixel window around the order center along the cross-dispersion direction (5 pixels above and 5 pixels below the order center). Normalization is done by dividing the flux value in each pixel of the window by the sum of fluxes in the window itself, ensuring the flux sum in each window is equal to 1. We chose the dimension of the window to be $10$ pixels, to include at least $95\%$ of the flux along the cross-dispersion direction.
\item We extract the 1D spectrum in each order using the optimal extraction described in \citep{horne1986optimal}. 

\item After computing the flux values for each order and converting the \(X\) positions into wavelength values, the next step consists of merging different orders to create 
the merged spectrum. Since each order shares some wavelengths with the next order, a method is needed to merge them seamlessly.\\
Firstly, I identify the overlapping region between consecutive orders and apply cubic splines to each spectrum in this region, considering the spectrum's wavelength, flux, and flux error. Subsequently, a common wavelength grid is generated within the overlapping region, and a weighted mean of the flux in the two consecutive orders is calculated. The weights for the weighted mean are determined based on the inverse of flux errors.


\end{enumerate}
An extracted 1D spectrum for an off-axis exposure of 1 hour and the corresponding signal-to-noise ratio are shown in Figure \ref{12}.

\subsection{Identify planet signal}



\begin{figure}[ht]
\centering
\includegraphics[width=\textwidth]{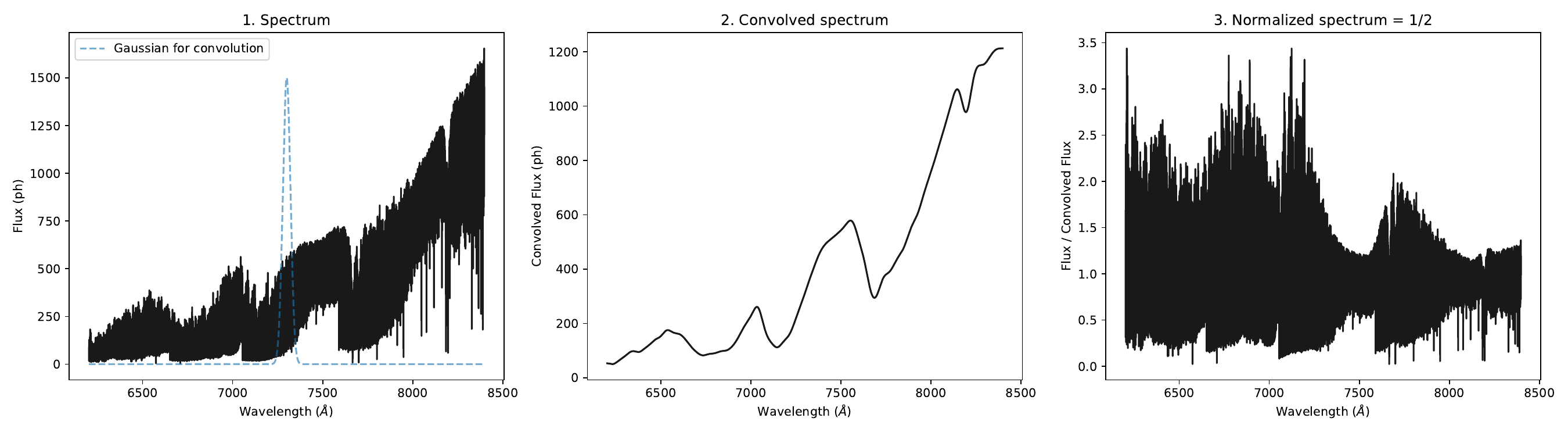}
\caption{Gaussian convolution and normalization of a spectrum to remove the low frequencies that induce spectral variations}
    \label{8}
\end{figure}

In this paper, we present a technique for detecting the planet and determining the inclination of its orbit, $i$, by disregarding the planet's atmospheric composition and phase function. Additionally, in the analyses that follow, we exclude points 5 and 6 of Section \ref{sec3}, namely the presented analyses were conducted without using Pyechelle, but only the simulated input spectra. The analyses involving the generated and extracted spectra from Pyechelle are ongoing, and we plan to publish those results in the near future.\\\\
\noindent Both the star and the planet are influenced by coupling functions that depend on their positions in the sky and the wavelength. These functions introduce flux variations that differ from spectrum to spectrum. The positions of astronomical objects can be affected by atmospheric turbulence or instrument-pointing instability, making it impossible to precisely predetermine these curves and predict their changes during observations.\\
To reduce the impact of flux variation in the spectra, which is mainly characterized by low-frequency variations, we normalize the spectra. First, we compute the convolution between each spectrum and a Gaussian with a Full Width at Half Maximum (FWHM) of $15$ nm. We then use this result to normalize the spectra. Figure \ref{8} shows an example of such normalization, representing a 1-hour exposure in an off-axis spaxel.
\\
\noindent Let's denote the extracted $99$ spectra of the off-axis spaxel where the planet is located (assuming prior knowledge of its position) as $F_1(\lambda)$, and $F_2(\lambda)$ as the spectra of the star alone placed at the center of the off-axis spaxel in a short exposure. 

\begin{figure}[ht]
\centering
\includegraphics[width=0.7\textwidth]{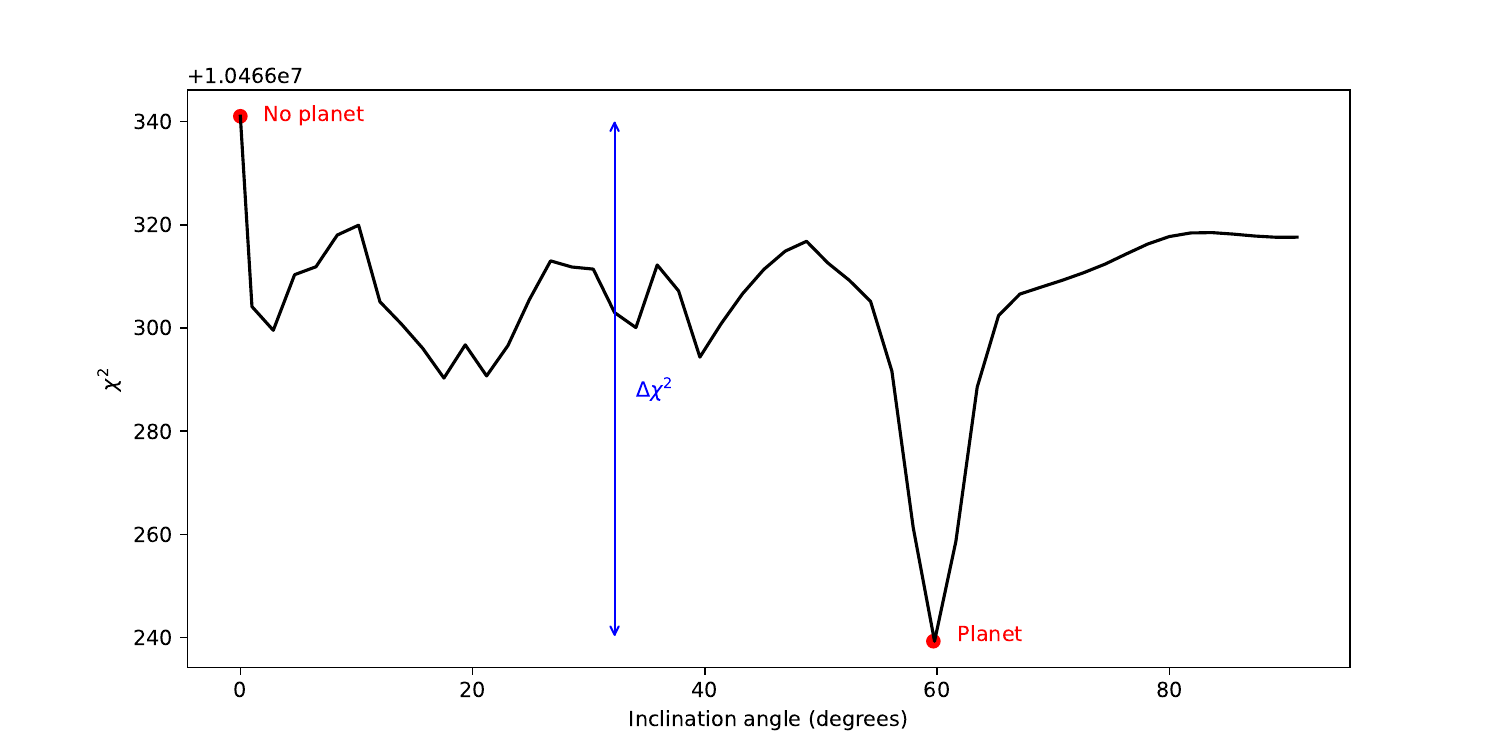}
\caption{Computed $\chi^2$ in the off-axis spaxel. The point at $i=0^{\circ}$ corresponds to a model that does not include a planet. To validate the detectability of the planet we use the $\Delta \chi^2$ shown in blue.}
    \label{10}
\end{figure}

\noindent We fitted the off-axis normalized flux with this model:
\begin{equation}
  \text{Model:}\quad   F_{1,\text{norm},j}=\beta F_{2,\text{norm},j}(\lambda)+(1-\beta)F_{2,\text{norm},j}(\lambda,RV_j)
  \label{model}
\end{equation}
Here, $j$ refers to the $j^{\text{th}}$ exposure, ranging from $0$ to $99$. 
Again, we make no assumptions about the planet's composition and so we model the planet by assuming it is equal to the star's halo multiplied by a constant.
The total number of free parameters in the fit is $2$: $\beta$ and $i$, where the latter constrains all the $RV_j$. Instead of performing a normal fit, we manually mapped the space of different $i$ values. Specifically, we selected $50$ different $i$ values between $0$ and $90^{\circ}$, created the corresponding list of $RV_j$ values, and used the least squares method to calculate the chi-squared value for each exposure to evaluate the fit:
\begin{equation}
\chi^2(i)=\sum_{j=0}^{99}\left(\frac{F_{1,\text{norm},j}-F_{1,\text{norm},j, \text{model}}}{\sigma_{F_{1,\text{norm},j}}}\right)^2
\label{chi2}
\end{equation}
Where the $\sigma_{F_{1,\text{norm}}}$ are calculated starting from the $\sigma_{F_{2,\text{norm}}}$.
\noindent In fact, the off-axis planet signal is much more noise-affected than the off-axis star signal and this could also impact the $\chi^2$ by worsening it. Therefore to compute $\sigma_{F_{1,\text{norm}}}$ we first took the ratio $F_{1}/F_{2}$ and fitted a polynomial of $3^{\text{rd}}$ degree on it, which we call $p(\lambda)$. Then we computed $\sigma_{F_{1}}=\sqrt{F_{2}\cdot p(\lambda)+\text{RON}^2}$ and divided it by the same convolved spectrum used to normalize $F_{1}$ to obtain $\sigma_{F_{1,\text{norm}}}$.\\
\noindent The result of the chi-squared map as a function of the inclination angle is shown by the black curve in Figure \ref{10}.\\ 
\noindent By examining Figure \ref{10}, we can observe the following remarks:\hfill

\begin{itemize}
\item The mean value of the $\chi^2$ is in agreement with the theoretical values. In fact, following the central limit theorem \citep{fisher1922mathematical}, a chi-squared distribution with a high number of $p$ degrees of freedom converges to the normal distribution of mean $p$ and variance $2p$. In our case, for the first model, we have $p=99n-2$, where $n$ is the number of data points (wavelength grid length) 
 and correspond to about $105'000$. Then, theoretically, the distribution of the chi-squared for our model should follow a normal distribution $N(10.4606\cdot10^6,0.0046\cdot10^6)$, which is compatible at better than $2\sigma$ with the black curve of Figure \ref{10}.
\item When $i=0$, $RV_j=0$, indicating that the models do not include the presence of a planet, as shown in the equation \ref{model}. It is therefore explanatory the fact that the first point of the chi-squared distribution is higher than the normal trend. This can be translated by saying that by assuming that in the off-axis there is only the halo of the star gives a worse result compared to a model where we include the planet presence.
    
\end{itemize}
To validate the detectability of the planet we used the Bayesian information criterion (BIC)  \citep{schwarz1978estimating}.
The BIC is a criterion for model selection among a finite set of models. It is based on the likelihood function and is penalized for the number of parameters in the model, thus balancing model fit and complexity. The BIC is defined as:
\[
\text{BIC} = k \ln(n) - 2 \ln(\hat{L})=k \ln(n)+\chi^2
\]
\noindent where \( k \) is the number of parameters in the model, \( n \) is the number of data points, and \( \hat{L} \) is the maximized value of the likelihood function of the model.
\noindent To determine whether the assumption that there is a planet in the data is statistically robust, we calculate the BIC corresponding to $i=60^{\circ}$ and the BIC at $i=0^{\circ}$, which again corresponds to a model without the presence of the planet. The model with the lowest BIC is preferred because it represents the best trade-off between goodness of fit and model complexity. A common rule of thumb is that if $\Delta$BIC $>$ 10, then the evidence for the model with the lower BIC value is significantly stronger compared to the other model. Conversely, if $\Delta$BIC $<$ 2, the evidence is considered weak, and the models are considered equivalent in terms of explanatory power.\\
By performing the same simulation $10$ times, with different random photon noise addition, we obtained that $\Delta$BIC = BIC$_{0^\circ}$ - BIC$_{60^\circ}$ = $-ln(n) +\chi^2_{0^\circ}-\chi^2_{60^\circ} > 40 $ 
This provides strong evidence that the model including the planet statistically fits the data better.\\
To determine $i$, we expanded the dataset by adding additional data points of $i$ ranging between 45 and 75. We then recalculated the $\chi^2$ and performed a Gaussian fit on the extended data. The fitted value of $i$ is $59.88^{\circ} \pm 0.29^{\circ}$, indicating a high level of accuracy and consistency with the true value $i=60^{\circ}$.

\section{Conclusion and perspectives}
In this proceeding, we presented a simulator for the future RISTRETTO spectrograph, which includes the optical model of the instrument and its coupling functions. Our primary objective was to validate the presence of Proxima b and determine its orbital inclination.\\
We employed a statistical analysis using the Bayesian Information Criterion (BIC) to evaluate whether including the planet in our model was justified. Our simulations indicate that Proxima b can be detected and its inclination angle ($i$) determined with 99 hours of exposure time.
For future work, we plan to repeat the analysis using spectra extracted from Pyechelle and to explore more robust statistical approaches, such as Bayesian Evidence with the nested sampling algorithm.\\
\noindent Furthermore, 
it would be possible to consider incorporating the planet's atmospheric composition and phase angle into the model to evaluate whether these additions yield a statistically more significant fit compared to the current model and we could extend the simulations to other planetary systems or astronomical objects to broaden the applicability of our findings. This latter would allow us to assess the versatility and robustness of the RISTRETTO spectrograph and simulator in different observational scenarios.

\renewcommand{\bibfont}{\normalsize} 
\bibliography{bibliography} 


\end{document}